\documentstyle[12pt]{article}
\newcommand{\nsect}{\setcounter{equation}{0}
\def\theequation{\thesection.\arabic{equation}}\section}
\newcommand{\nappend}{\setcounter{equation}{0}
\def\theequation{\rm{A}.\arabic{equation}}\section*}

\newcommand{\newc}{\newcommand}
\newc{\del}{\partial}
\newc{\beq}{\begin{equation}}
\newc{\eeq}{\end{equation}}
\newc{\barr}{\begin{eqnarray}}
\newc{\earr}{\end{eqnarray}}
\newc{\ra}{\rightarrow}
\newc{\lam}{\lambda}
\newc{\eps}{\epsilon}
\newc{\half}{\frac{1}{2}}
\newc{\third}{\frac{1}{3}}
\newc{\fourth}{\frac{1}{4}}
\newc{\eighth}{\frac{1}{8}}
\newc{\gev}{\,GeV}
\newc{\lra}{\leftrightarrow}
\newc{\Dslash}{\not\!\! D}
\newc{\sg}{{\cal G}}
\newc{\etal}{{\it et al.}\ }
\newc{\Hbar}{{\bar H}}
\newc{\hhbar}{{\overline h}}
\newc{\Ubar}{{\bar U}}
\newc{\Dbar}{{\bar D}}
\newc{\Ebar}{{\bar E}}
\newc{\eg}{{\it e.g.}\ }
\newc{\ie}{{\it i.e.}\ }
\newc{\nonum}{\nonumber}
\newc{\kap}{\kappa}
\newc{\Dt}{\frac{d}{dt}}
\newc{\rpv}{$\not\!\!R_p$}
\newc{\bpv}{$\not\!\!B_p$}
\newc{\mpl}{$M_{Pl}$\ }
\newc{\mx}{$M_X$\ }
\newc{\tev}{\,TeV}
\newc{\sect}[1]{\ref{sec:#1}}
\newc{\nonr}{\nonumber}
\newc{\eq}[1]{(\ref{eq:#1})}
\newc{\eqs}[2]{(\ref{eq:#1},\ref{eq:#2})}
\newc{\lab}[1]{\label{eq:#1}}
\newc{\Lam}{{\bf\Large \Lambda}}
\newc{\lame}[1]{{\Lam}_{E^{#1}}}
\newc{\lamd}[1]{{\Lam}_{D^{#1}}}
\newc{\lamu}[1]{{\Lam}_{U^{#1}}}
\newc{\lamet}[1]{{\Lam}_{E^{#1}}^T}
\newc{\lamdt}[1]{{\Lam}_{D^{#1}}^T}
\newc{\lamut}[1]{{\Lam}_{U^{#1}}^T}
\newc{\lames}[1]{{\Lam}_{E^{#1}}^*}
\newc{\lamds}[1]{{\Lam}_{D^{#1}}^*}
\newc{\lamus}[1]{{\Lam}_{U^{#1}}^*}
\newc{\lamed}[1]{{\Lam}_{E^{#1}}^\dagg}
\newc{\lamdd}[1]{{\Lam}_{D^{#1}}^\dagg}
\newc{\lamud}[1]{{\Lam}_{U^{#1}}^\dagg}
\newc{\Y}{{\bf Y}}
\newc{\ye}{{\Y}_E}
\newc{\yd}{{\Y}_D}
\newc{\yu}{{\Y}_U}
\newc{\yes}{{\Y}_E^*}
\newc{\yds}{{\Y}_D^*}
\newc{\yus}{{\Y}_U^*}
\newc{\yet}{{\Y}_E^T}
\newc{\ydt}{{\Y}_D^T}
\newc{\yut}{{\Y}_U^T}
\newc{\yed}{{\Y}_E^\dagg}
\newc{\ydd}{{\Y}_D^\dagg}
\newc{\yud}{{\Y}_U^\dagg}
\newc{\dagg}{\dagger}
\newc{\lp}{\left(}
\newc{\rp}{\right)}
\newc{\inv}{\frac{1}{16\pi^2}}
\newc{\invsq}{\frac{1}{(16\pi^2)^2}}
\newc{\ggam}[2]{\Gamma_{#1}^{#2}}
\newc{\yukgam}[2]{\inv \gamma_{#1}^{(1){#2}}+\invsq\gamma_{{#1}}^{(2){#2}}}
\newc{\susyunif}{ohman,nirpaul,marcelacarlos,susyunif}
\newc{\lsim}{\stackrel{<}{\sim}}
\newc{\gsim}{\stackrel{>}{\sim}}
\begin{document}

\title{2-Loop Supersymmetric Renormalization Group Equations
Including R-parity Violation and Aspects of Unification}
\author{Herbi Dreiner$^{1,2}$\thanks{Permanent address as of February 5$^{th}
$, 1996, Rutherford Appleton Laboratory, Chilton, Didcot, Oxon, OX11 0QX, UK.}
 \ and Heath Pois$^{3,4}$}
\date{$^1${\small Theoretische Physik, ETH-H\"onggerberg, CH-8093 Z\"urich,
Switzerland}\\
$^2${\small ITP, UC Santa Barbara, Santa Barbara, CA 93106-4030, USA}\\
$^3$ {\small Therma-Wave, 47320 Mission Falls Ct, Freemont, CA 94539, USA \\
$^4$ Davis Institute for High Energy Physics, Dept. of Physics,
U.C. Davis, CA 95616, USA }}
\maketitle

\vspace{-10.3cm}
\hfill\parbox{8cm}{\raggedleft \today \\ ETH-TH/95-30\\
NSF-ITP-95-155}
\vspace{8.5cm}

\begin{abstract}
\noindent We present the complete 2-loop renormalization group equations of the
supersymmetric standard model. We thus explicitly include the full set of $R
$-parity violating couplings, including $\kappa_iL_iH_2$. We use these
equations to do a first study of $(a)$ gauge coupling unification, $(b)$
bottom-tau unification, $(c)$ the fixpoint structure of the top quark Yukawa
coupling, and (d) two-loop bounds from perturbative unification. We find
significant shifts which can be larger than the effect from the top quark
Yukawa coupling. The value of $\alpha_3(M_Z)$ can change by $\pm5\%$. The
$\tan\beta$ region for bottom-tau unification and for the top quark IR quasi
fixed point structure is significantly increased. For heavy scalar fermion
masses ${\cal{O}}(1\tev)$ the limits on the $\Delta L\not=0$ operators from
perturbative unification are competitive with the indirect laboratory bounds.
The two-loop correction to the bound on the $\Delta B\not=0$ opertor is $+15\%
$.
\end{abstract}

\nsect{Introduction}
The most compelling indication for supersymmetry is the unification of the
gauge coupling constants. This has been thoroughly investigated in the
literature \cite{ohman}-\cite{susyunif}, mainly in the context of the minimal
supersymmetric standard model (MSSM). In all cases it was assumed that R-parity
\beq
R_p=(-)^{3B+L+2S}
\eeq
is conserved.\footnote{B: Baryon number, L: Lepton number, S: Spin.} However,
$R_p$ is imposed {\it ad hoc}: there is no experimental or theoretical
justification. $R_p$ is sufficient but not necessary to guarantee the stability
of the proton. Other discrete \cite{rossiban} or gauge symmetries are equally
possible \cite{aliherbi1,aliherbi2}. Strict cosmological bounds based on
GUT-scale baryogenesis have been proposed \cite{brucefischler} but these have
been shown to be strongly model dependent \cite{sphalgraham}.
The best motivation for R-parity is that
it offers a good candidate for as yet undetected cold dark matter. However,
since this has not directly been observed R-parity violation (\rpv) should be
considered on equal footing with conserved R-parity in supersymmetric model
building and especially in collider searches \cite{grahamherbi,bgh}.

Throughout this paper we allow for \rpv. Thus we consider the additional terms
in the superpotential
\beq
W_{\not R_p}=\lam_{ijk}L_iL_j\Ebar_k+\lam'_{ijk}L_iQ_j\Dbar_k
+\lam''_{ijk}\Ubar_i\Dbar_j\Dbar_k+\kappa_iL_iH_2.
\lab{rpsuper}
\eeq
We have used superfield notation and the fields have the $G_{SM}=SU(3)_c
\times SU(2)_L\times U(1)_Y$ quantum numbers
\barr
L:&(1,2,-\half),\quad {\bar E}:&(1,1,1),\qquad\, Q:\,(3,2,\frac{1}{6}),\quad
{\bar U}:\,(3,1,\frac{2}{3}),\nonr\\ {\bar D}:&(3,1,-\frac{1}{3}),\quad
H_1:&(1,2,-\half),\quad  H_2:\,(1,2,\half).
\lab{fields}
\earr
$i,j,k$ are generation indices. $\lam_{ijk}$ is anti-symmetric in the first
two indices and $\lam''_{ijk}$ is anti-symmetric in the last two.

Models for unification are typically constructed at very high energies, such as
grand unified theories (GUTs). Such models can predict the absolute or relative
size of parameters at the unification scale. In order to compare the
predictions from such models with the low-energy data we must employ the
renormalization group (RG). Recently there have been several studies of the
RG properties of the \rpv-Yukawa couplings at one-loop
\cite{rogervern}-\cite{ralf}. These have been used to place first (weak)
bounds on several of the higher generation operators via unitarity constraints
\cite{prob}-\cite{marcprob}. They have also been used to compare \rpv-GUT
predictions with the low-energy data \cite{smirnov,ralf}.

The main interest in the evolution of the gauge coupling constants is whether
they unify at a high scale, \eg grand-unified or string. The \rpv-Yukawa
couplings have to date been neglected in this context, mainly because the
effect was expected to be small. However, as we show the effect can be larger
than the contribution from the top quark Yukawa, $\lam_t$, which can not be
neglected. The main reason is that the higher generation couplings are only
weakly bound if at all \cite{marcgoi,prob,gautam}. Also, the bounds as
presented are usually scaled with the mass dependent factor $({\tilde m}/100
\gev)$ and for SUSY masses of order $1\tev$ the bound is typically weaker than
the bound from perturbative unification. The strictest bounds for a mass of $0.
1(1)\tev$ for the couplings that we will be considering are
\barr
\lam_{323}<0.09\,(0.9),\quad\lam'_{333}<0.45\,(4.5)^*,\quad\lam''_{323}<
1.25. \lab{bounds}
\earr
At $1\tev$ the bound on $\lam_{323}$ \cite{bgh} is almost identical to the
perturbative limit obtained below in Section 3. The bound on $\lam'_{333}$
\cite{gautam} at $1\tev$, indicated by an asterisk is obtained by scaling and
as such is meaningless since perturbation theory breaks down for such large
values. The appropriate bound is thus the perturbative limit. The bound on $
\lam''_{323}$ is the bound from perturbative unification \cite{marcgoi,prob}.
We shall thus explore all three couplings to the pertrubative limit.

We propose to investigate the effect of \rpv\ on the unification of the gauge
coupling constants. The Yukawa couplings only enter at the 2-loop level. We
thus extend previous work and present the full two-loop renormalization
group equations for \rpv\  in Section 2.
We then focus on the applications of these equations in Section 3. We
investigate the unification of the gauge couplings, bottom-tau unification, and
the fixed-point structure of $\lam_t$. We shall perform three case studies, one
dominant operator from $LL\Ebar,\,LQ\Dbar,$ and $\Ubar\Dbar\Dbar$ respectively,
to illustrate the possible \rpv\ effects. Given the laboratory bounds
\eq{bounds},
the third generation couplings can be the largest and we shall focus on them
for maximum effect. However, it is {\it not} necessarily the theoretical
expectation that the third generation couplings dominate. If the Standard Model
Yukawas and the \rpv-Yukawas have a common physical origin, a symmetry, then
we expect this symmetry to distinguish between a Higgs superfield and the
lepton
doublet superfields. If the hierarchy in the Yukawa couplings is determined by
the structure of the broken symmetry we would expect the resulting hierarchy to
have a different flavour structure for the Higgs couplings then for the purely
matter-field couplings \cite{aliherbi1,aliherbi2}.

When studying the RG evolution of the gauge couplings the central interest is
unification. Does it make sense to discuss \rpv\  in the context of
unification? In order to avoid rapid proton decay we must require a symmetry
which treats quarks and leptons differently. This is counter to any expectation
{}from GUTs where they are in common multiplets. All the same, several
supersymmetric grand unified models have been constructed with a low-energy
R-parity violating superpotential \cite{hallsuz,hallbrahm,smirnov,ralf}. These
models transfer the extreme mass splitting in the Higgs sector to an
asymmetry in the quark-lepton multiplet via $LH_2$ mixing. In order to keep
this
mixing small it is generated through a broken symmetry. They have no further
fine-tuning and the proton decay rate is consistent with experiment. It is
thus highly relevant to consider the effects of \rpv\  on gauge coupling
unification.

In string theories there is no preference for grand unification, and
unification may very well be obtained with a non-simple group such as $G_{SM}$.
As discussed in detail in Ref.\cite{bentohallross}, there is then also no
preference for $R_p$ over for example baryon parity $B_p=(-)^{3B+2S}$. The
authors obtain several models with either $R_p$ or \rpv. The main question
in string unification is whether the correct value of $sin^2\theta_W$ can be
obtained. Again, in this context the RGEs for the \rpv-Yukawa couplings must be
considered.

In GUTs the prediction $m_b(M_U)=m_\tau(M_U)$ has been very successful
\cite{btauguys,fixedpoint1}. In the MSSM if one requires the Yukawa couplings
to unify this greatly reduces the allowed region of the
(supersymmetric) parameters. In particular one obtains a strict relation
between the running top mass $m_t(m_t)$ and the ratio of the vacuum expectation
values $\tan\beta$ \cite{ohman,marcelacarlos}. Given the observed top quark
mass \cite{cdf} this results in a prediction for $\tan\beta$. How general is
this prediction? The third generation \rpv-couplings enter the evolution of
$m_t,\,m_b$, and $m_\tau$ at one loop and can thus have a large effect. Thus
if we allow for \rpv\ we expect the strict predictions of the MSSM to be
modified. In Section 3.2 we shall analyze this affect and determine a new
$\tan\beta$ solution for bottom-tau unification.

With the recent discovery of the top quark we have determined all the fermion
masses in the Standard Model. It seems that the top quark mass is special.
There has recently been much work to predict the fermion masses at the weak
scale from a simple symmetry structure at the unification scale
\cite{fermmassguys,fixedpoint1}. It is possible that the fermion mass
structure is determined by a broken symmetry \cite{massibross} where only the
top-quark
Yukawa coupling is allowed by the symmetry at tree-level. It's value is put in
by hand, presumably of order one. The other couplings are then determined
dynamically through the symmetry breaking model. Given such a model, we would
then still require a prediction for the top-quark Yukawa coupling. An
intriguing possibility is that this Yukawa coupling is given by an
infra-red (quasi) fix point \cite{ir}. The low-energy value then depends
only very weakly on the high-energy initial value; the exact opposite of a
fine-tuning problem. In supersymmetric GUTs with bottom-tau unification one
typically requires large values of $\lam_t\sim1$ close to the IR quasi
fixed-point. This has been studied in detail in Refs.\cite{fixedpoint1,ohman,
marcelacarlos,fixedpoint2,topfix}. In the MSSM this is only attained for a
small range of parameters, in particular $\tan\beta$. We investigate the
effect of the \rpv-couplings on this scenario in Section 3.4. Similar to the
case of bottom-tau unification, in Section 3.3 we find a new $\tan\beta$
solution with fixed-point structure.

\nsect{Renormalization Group Equations}
We apply the work of Martin and Vaughn (MV) \cite{mv} to the superpotential
\eq{superpot}. We shall closely follow their notation. At this point we retain
the coupling $\kappa_iL_iH_2$ to be most general and defer a discussion to
Section 2.4. We denote the $G_{SM}$ gauge couplings by
\beq
g_3,\quad g_2,\quad g_1.
\eeq
The chiral superfields are given in (\ref{eq:fields}). In Appendix A we have
collected several useful group theoretical formulas pertaining to $G_{SM}$ and
the above field content. Here we mention that for $U(1)_Y$ we use the
normalization as in grand unified theories and thus use $g_1\rightarrow g_Y$.
See the appendix for more details. We define our notation for the Yukawa
couplings via the superpotential including all \rpv\ terms.
\barr
W&=&
(\ye)_{ij} L_iH_1{\bar E}_j +
(\yd)_{ij} Q_iH_1{\bar D}_j +
(\yu)_{ij} Q_iH_2{\bar U}_j\nonr \\
&&+(\lame{k})_{ij} L_iL_j{\bar E}_k +
(\lamd{k})_{ij} L_iQ_j{\bar D}_k +
(\lamu{i})_{jk} {\bar U}_i{\bar D}_j{\bar D}_k
\nonr\\&& +\mu H_1H_2+\kappa^iL_iH_2.
\lab{superpot}
\earr
We have introduced the twelve $3\times3$ matrices
\beq
\ye,\quad \yd,\quad \yu,\quad \lame{k},\quad \lamd{k},\quad \lamu{i},
\lab{matrices}
\eeq
for all the Yukawa couplings. This implies the following conventions
\barr
Y_{L_iL_j{\bar E}_k}&=&Y_{L_i{\bar E}_kL_j}=Y_{{\bar E}_kL_iL_j}
= \left({\bf\Lambda}_{E^k}\right)_{ij},\\
Y_{L_jL_i{\bar E}_k}&=&Y_{L_j{\bar E}_kL_i}=Y_{{\bar E}_kL_jL_i}
=\left({\bf\Lambda}_{E^k}\right)_{ji}=-\left({\bf\Lambda}_{E^k}\right)_{ij},
\\
Y_{{\bar U}_i{\bar D}_j{\bar D}_k}&=&Y_{{\bar D}_j{\bar U}_i{\bar D}_k}=
Y_{{\bar D}_j{\bar D}_k{\bar U}_i}= \left({\bf\Lambda}_{U^i}\right)_{jk},\\
Y_{{\bar U}_i{\bar D}_k{\bar D}_j}&=&Y_{{\bar D}_k{\bar U}_i{\bar D}_j}=
Y_{{\bar D}_k{\bar D}_j{\bar U}_i}= \left({\bf\Lambda}_{U^i}\right)_{kj}=-
\left({\bf\Lambda}_{U^i}\right)_{jk}.
\earr
We now in turn study the dimensionless couplings and then briefly also discuss
the mass terms $\mu,\kappa_i$. We do not here consider the soft-breaking terms.

\subsection{Gauge Couplings}
The renormalization group equations for the gauge couplings are given by
\beq
\frac{d}{dt}g_a = \frac{g_a^3}{16\pi^2} B_a^{(1)}
+\frac{ g_a^3}{(16\pi^2)^2} \left[ \sum_{b=1}^3 B_{ab}^{(2)} g_b^2
-\sum_{x=u,d,e} \left(C_a^x Tr(\Y_x^\dagger \Y_x)+
A_a^x \sum_{i=1}^3Tr(\Lam_{x_i}^\dagger \Lam_{x_i})\right)
\right].
\lab{gaugerg}
\eeq
The coefficients $B_a,\,B_{ab},$ and $C_a^x$ have been given previously
\cite{bjorkjones} and for completeness we present them in the appendix. The
\rpv-effects on the running of the gauge couplings are new, we obtain
\beq
A_{a}^{u,d,e}=\left( \begin{array}{ccc}
12/5 & 14/5 & 9/5 \\
0 & 6 & 1 \\
4 & 4 & 0
\end{array}
\right).
\eeq
This completes the equations for the running of the gauge coupling constants
at two-loop.

\subsection{Yukawa Couplings}
In general the renormalization group equations for the Yukawa
couplings are given by \cite{mv}
\beq
\frac{d}{dt} Y^{ijk}=Y^{ijp}\left[\frac{1}{16\pi^2}\gamma_p^{(1)k}
+\frac{1}{(16\pi^2)^2}\gamma_p^{(2)k}\right]+(k\lra i)+(k\lra j),
\lab{rgyukawa}
\eeq
and the one- and two-loop anomalous dimensions are
\barr
\gamma_i^{(1)j}&=& \half Y_{ipq}Y^{jpq}-2\delta_i^j\sum_ag_a^2C_a(i),\\
\gamma_i^{(2)j}&=&-\half Y_{imn}Y^{npq}Y_{pqr}Y^{mrj}+Y_{ipq}Y^{jpq}
\sum_a g_a^2[2C_a(p)-C_a(i)]\nonr \\&&+2\delta_i^j\sum_ag_a^2
\left[g_a^2C_a(i)
S_a(R)+2\sum_bC_a(i)C_b(i)-3g_a^2C_a(i)C(G_a)\right]. \lab{gamtwo}
\earr
See the appendix for the definition of the group theoretical quantities
$C_a(f),\,S_a(R),$ and $C(G_a)$.
We now first give the explicit version of Eq.\eq{rgyukawa} for the
matrices \eq{matrices} in terms of the anomalous dimensions, and
then we present the explicit forms for $\gamma^{(1)f_j}_{f_i},$ and
$\gamma^{(2)f_j}_{f_i}$.

\subsubsection{RG-Equations}
The RGEs Yukawa couplings are given by
\barr
\Dt (\yes)_{ik} &=& (\yes)_{ip}\ggam{E_p}{E_k}\lab{ye}
+(\yes)_{ik}\ggam{H_1}{H_1}
+(\lames{k})_{ip}\ggam{L_p}{H_1}
+(\yes)_{pk}\ggam{L_p}{L_i},\\
\Dt (\yds)_{ik} &=& (\yds)_{ip}\ggam{D_p}{D_k}
  +(\yds)_{ik}\ggam{H_1}{H_1}
  +(\lamds{k})_{pi}\ggam{L_p}{H_1}
  +(\yds)_{pk}\ggam{Q_p}{Q_i},\lab{yd}\\
\Dt (\yus)_{ik} &=& (\yus)_{ip}\ggam{U_p}{U_k}\lab{yu}
+(\yus)_{ik}\ggam{H_2}{H_2}
+(\yus)_{pk}\ggam{Q_p}{Q_k}, \lab{rgyup}
\\
\Dt (\lames{k})_{ij} &=& (\lames{p})_{ij}\ggam{E_p}{E_k}\lab{lame}
+(\lames{k})_{ip}\ggam{L_p}{L_j}
+(\yes)_{ik}\ggam{H_1}{L_j}
+(\lames{k})_{pj}\ggam{L_p}{L_i}+(\yes)_{jk}\ggam{H_1}{L_i},\\
\Dt (\lamds{k})_{ij} &=& (\lamds{p})_{ij}\ggam{D_p}{D_k}\lab{lamd}
+(\lamds{k})_{ip}\ggam{Q_p}{Q_j}
+(\lamds{k})_{pj}\ggam{L_p}{L_j}
+(\yds)_{jk}\ggam{H_1}{L_i},\\
\Dt (\lamus{i})_{jk} &=& (\lamus{i})_{jp}\ggam{D_p}{D_k} \lab{lamu}
+(\lamus{i})_{pk}\ggam{D_p}{D_j}
+(\lamus{p})_{jk}\ggam{U_p}{U_i}.
\earr
At two-loop the anomalous dimensions are given by
\beq
\ggam{f_i}{f_j}=\yukgam{f_i}{f_j}.
\eeq
As we discuss below in section 2.4 in many cases of interest the anomalous
dimension $\Gamma_{L_i}^{H_1}$ vanishes.

\subsubsection{Anomalous Dimensions}
The one-loop anomalous dimensions are given by
\barr
\gamma^{(1)L_j}_{L_i} &=&\left(\ye \ye^\dagg \right)_{ij}
+(\lame{q}\lame{q}^\dagg)_{ij} +3 (\lamd{q}\lamd{q}^\dagg)_{ij}
-\delta_i^j(\frac{3}{10}g_Y^2+\frac{3}{2}g_2^2), \lab{gamll1}\\
\gamma^{(1)E_j}_{E_i} &=& 2 \left(\ye^\dagg \ye \right)_{ji}
+ Tr(\lame{i}\lame{j}^\dagg) -\delta_i^j(\frac{6}{5}g_Y^2),\lab{gamee1}\\
\gamma^{(1)Q_j}_{Q_i} &=& \left(\yd \yd^\dagg \right)_{ij}
+ \left(\yu \yu^\dagg \right)_{ij}
+ (\lamd{q}^\dagg\lamd{q})_{ji}
-\delta_i^j(\frac{1}{30}g_Y^2+\frac{3}{2}g_2^2+\frac{8}{3}g_3^2),\\
\gamma^{(1)D_j}_{D_i} &=& 2 \left(\yd^\dagg \yd \right)_{ji}
+2 Tr(\lamd{i}\lamd{j}^\dagg)
+2 (\lamu{q}\lamu{q}^\dagg)_{ij}
-\delta_i^j(\frac{2}{15}g_Y^2+\frac{8}{3}g_3^2)),\\
\gamma^{(1)U_j}_{U_i} &=& 2\left(\yu^\dagg \yu \right)_{ij}
+ Tr(\lamu{i}\lamu{j}^\dagg)
-\delta_i^j(\frac{8}{15}g_Y^2+\frac{8}{3}g_3^2)),\\
\gamma^{(1)H_1}_{H_1} &=& Tr\left(3\yd\yd^\dagg+\ye\ye^\dagg \right)
-(\frac{3}{10}g_Y^2+\frac{3}{2}g_2^2),\\
\gamma^{(1)H_2}_{H_2} &=& 3 Tr\left( \yu\yu^\dagg\right)
-(\frac{3}{10}g_Y^2+\frac{3}{2}g_2^2),\\
\gamma^{(1)H_1}_{L_i} &=& 3 (\lamd{q}\yd^*)_{iq}
+ (\lame{q}\ye^*)_{iq}.
\earr
For the two-loop anomalous dimensions we write
\beq
\gamma_{f_i}^{(2)f_j}=\left(\gamma_{f_i}^{(2)f_j}\right)_{yukawa}+
\left(\gamma_{f_i}^{(2)f_j}\right)_{g-y}+
\left(\gamma_{f_i}^{(2)f_j}\right)_{gauge}.
\eeq
These correspond respectively to the three terms of \eq{gamtwo}. These are
given explicitly below. The pure gauge two-loop anomalous dimensions are
given by
\barr
\left(\gamma^{(2){L_j}}_{L_i}\right)_{gauge} &=& \delta_i^j
(\frac{15}{4}g_2^4+\frac{207}{100}g_Y^4+\frac{9}{10}g_2^2g_Y^2),\\
\left(\gamma^{(2){E_j}}_{E_i}\right)_{gauge} &=& \delta_i^j\frac{234}{25}
g_Y^4,\\
\left(\gamma^{(2){Q_j}}_{Q_i}\right)_{gauge}&=&\delta_i^j (
-\frac{8}{9}g_3^4+\frac{15}{4}g_2^4+\frac{199}{900}g_Y^4 + 8g_3^2 g_2^2 +
\frac{8}{45} g_3^2g_Y^2 + \frac{1}{10}g_2^2g_Y^2),\\
\left(\gamma^{(2){D_j}}_{D_i}\right)_{gauge} &=&\delta_i^j (
-\frac{8}{9}g_3^4+\frac{202}{225}g_Y^4 +\frac{32}{45} g_3^2g_Y^2),\\
\left(\gamma^{(2){U_j}}_{U_i}\right)_{gauge} &=&\delta_i^j (
-\frac{8}{9}g_3^4+\frac{856}{225}g_Y^4 +\frac{128}{45} g_3^2g_Y^2),\\
\left(\gamma^{(2){H_1}}_{H_1}\right)_{gauge} &=&
\left(\gamma^{(2){H_2}}_{H_2}\right)_{gauge} =
\left(\gamma^{(2){L_j}}_{L_i}\right)_{gauge}, \\
\left(\gamma^{(2){H_1}}_{L_i}\right)_{gauge} &=&0,
\earr
The mixed gauge-Yukawa two-loop anomalous dimensions are given by
\barr
\left(\gamma^{(2){L_j}}_{L_i}\right)_{g-y} &=& (16g_3^2-\frac{2}{5}g_Y^2)
\left(\lamd{q}\lamdd{q} \right)_{ij} + \frac{6}{5} g_Y^2 (\ye\ye^\dagg
+\lame{q}\lame{q}^\dagg)_{ij},
\\
\left(\gamma^{(2){E_j}}_{E_i}\right)_{g-y} &=&(6g_2^2-\frac{6}{5}g_Y^2)
(\ye^\dagg\ye)_{ji} +(3g_2^2-\frac{3}{5}g_Y^2) Tr(\lame{i}\lame{j}^\dagg),
\\
\left(\gamma^{(2){Q_j}}_{Q_i}\right)_{g-y} &=& \frac{2}{5}g_Y^2
[\left( \yd\yd^\dagg+2\yu\yu^\dagg\right)_{ij}
+\left(\lamd{q}^\dagg\lamd{q}\right)_{ji} ],
\\
\left(\gamma^{(2){D_j}}_{D_i}\right)_{g-y} &=&(\frac{16}{3}g_3^2+
\frac{16}{15}
g_Y^2)\left(\lamu{q}\lamu{q}^\dagg \right)_{ij}\nonr \\
&&+(6g_2^2+\frac{2}{5}g_Y^2)[\left(\yd\yd^\dagg\right)_{ij}+
Tr(\lamd{i}\lamd{j}^\dagg)],
\\
\left(\gamma^{(2){U_j}}_{U_i}\right)_{g-y} &=& (6g_2^2-\frac{2}{5}g_Y^2)
\left(\yu\yu^\dagg\right)_{ij}+
(\frac{8}{3}g_3^2-\frac{4}{15}g_Y^2)Tr(\lamu{i}\lamu{j}^\dagg),\\
\left(\gamma^{(2){H_1}}_{H_1}\right)_{g-y} &=& (16g_3^2-\frac{2}{5}g_Y^2)
Tr(\yd\yd^\dagg) + \frac{6}{5}g_Y^2 Tr(\ye\ye^\dagg), \\
\left(\gamma^{(2){H_2}}_{H_2}\right)_{g-y} &=& (16g_3^2+\frac{4}{5}g_Y^2)
Tr\left(\yu\yu^\dagg\right),\\
\left(\gamma^{(2){H_1}}_{L_i}\right)_{g-y} &=& (16g_3^2-\frac{2}{5}g_Y^2)
\left(\lamd{q}\yd^*\right)_{iq} +\frac{6}{5}g_Y^2\left(\lame{q}\ye^*
\right)_{iq}.
\earr
The pure Yukawa two-loop anomalous dimensions are given by
\barr
-\left(\gamma^{(2){L_j}}_{L_i}\right)_{yukawa} &=& 2\lp\ye\yed\ye\yed\rp_{ij}
+\lp\ye\yed\rp_{ij}Tr\lp\ye\yed+3\yd\ydd\rp\nonr \\
&+&\lp\ye\rp_{in}\lp\yed\rp_{rj}Tr\lp\lamed{n}\lame{r}\rp
+3\lp\ye\rp_{im}\lp\lamed{m}\lamd{q}\yds\rp_{jq} \nonr \\
&+&\lp\ye\rp_{im}\lp\yed\lame{q}
\lames{m}\rp_{qj}+\lp\lame{n}\lamed{r}\rp_{ij}Tr\lp\lamed{n}\lame{r}\rp \nonr
 \\
&+&2\lp\lame{n}\lames{r}\rp_{ij}\lp\yed\ye\rp_{nr}+\lp\lame{m}\yes\ye
\lames{m}
\rp_{ij}+\lp\lame{m}\lames{q}\ye\rp_{iq}\lp\yed\rp_{mj}\nonr \\
&+&3\lp\lame{m}\lamds{q}\yd\rp_{iq}\lp\yed\rp_{mj}+3\lp\lame{m}\lamds{q}
\lamdt{q}\lames{m}\rp_{ij}\nonr\\
&+&6\lp\lamd{n}\lamdd{r}\rp_{ij}\lp\lp\ydd\yd\rp_{nr}+\lp\lamus{q}\lamu{q}
\rp_{nr}+Tr\lp\lamdd{n}\lamd{r}\rp\rp\nonr\\
&+& 3\lp\lamd{m}\yds\ydt\lamdd{m}\rp_{ij}+3\lp\lamd{m}\yus\yut\lamdd{m}
\rp_{ij}
+\lp\lame{m}\lames{q}\lame{q}\lames{m}\rp_{ij} \nonr \\
&+&3\lp\lamd{m}\lamdd{q}\lamd{q}
\lamdd{m}\rp_{ij},\\
-\left(\gamma^{(2){E_j}}_{E_i}\right)_{yukawa} &=&
2\left(\ye^\dagg\ye
\right)_{ji}Tr\left(\ye^\dagg\ye+3\yd\yd^\dagg\right)+2 \lp\yed\ye
\yed\ye \rp_{ji}\nonumber \\
&+&2\lp\yed\lame{q}\lame{j}^\dagg\ye \rp_{qi}
+6\lp\ydd\lamd{q}^T\lame{j}^\dagg\ye\rp_{qi}+2\lp\yed\lame{q}^T
\lamed{q}\ye \rp_{ji}\\
&+&6\lp \yed\lamd{q}\lamdd{q}\ye\rp_{ji}
+ 2Tr\lp\lame{j}^*\ye\yed\lame{i}^T\rp+2\lp\yed\lame{i}\lames{q}\ye\rp_{jq}
\nonr\\
&+&6\lp\yed\lame{i}\lamds{q}\yd\rp_{jq}
 +2Tr\lp\lame{i}\lames{q}\lame{q}\lamed{j}\rp +6Tr\lp\lames{j}\lamd{q}
\lamdd{q}\lamet{i}\rp,\nonr
\\
-\left(\gamma^{(2){Q_j}}_{Q_i}\right)_{yukawa} &=&
2\lp\yd\ydd\yd\ydd\rp_{ij}
+\lp\yd\ydd\rp_{ij}Tr\lp\yed\ye+3\ydd\yd\rp \nonr \\
&+&2\lp\yu\yud\yu\yud\rp_{ij}+3\lp\yu\yud\rp_{ij}Tr\lp\yud\yu\rp\nonr \\
&+&2\lp\yd\lamus{q}\lamu{q}\ydd\rp_{ij}
+3\lp\lamdd{m}\lamd{q}\lamdd{q}\lamd{m}\rp_{ji} \nonr \\
&+& 2\lp\yd\rp_{in}\lp\ydd\rp_{rj} Tr\lp\lamdd{n}\lamd{r}\rp
+ \lp\lamdd{m}\ye\yed\lamd{m}\rp_{ji} \nonr \\
&+& \lp\lamdd{m}\lamet{q}\lamed{q}\lamd{m}\rp_{ji}
+\lp\yd\rp_{im}
\lp 3\ydd\lamdt{q}\lamds{m}
+\yed\lame{q}\lamds{m}\rp_{qj} \nonr \\
&+&\lp\yu\rp_{in}\lp\yud\rp_{rj}Tr\lp\lamud{n}\lamu{r}\rp \nonr \\
&+&2\lp\lamdd{r}\lamd{n}\rp_{ji}\lp\lp\ydd\yd+\lamus{q}\lamu{q}\rp_{nr}+
Tr\lp\lamdd{n}\lamd{r}\rp\rp\nonr \\
&+&\lp\ydd\rp_{mj}\lp3\lamdt{m}\lamds{q}\yd+\lamdt{m}\lames{q}\ye\rp_{iq},
\lab{gam2qq}\\
-\left(\gamma^{(2){D_j}}_{D_i}\right)_{yukawa} &=&
2\lp\ydd\yd\ydd\yd\rp_{ji}+2\lp\ydd\yu\yud\yd\rp_{ji}\nonr \\
&+&2\lp\ydd\yd\rp_{ji}Tr\lp\yed\ye+3\ydd\yd\rp
+2\lp\ydd\lamdt{q}\lamds{q}\yd\rp_{ji}\nonr \\
&+&2\lp\yed\lame{q}\lamds{j}\yd\rp_{qi}
+6\lp\ydd\lamdt{q}\lamds{j}\yd\rp_{qi}\nonr \\&+&
2 Tr\lp\lamdt{i}\lamds{j}\lp\yd\ydd+\yu\yud\rp+\lamd{i}\lamdd{q}\lamd{q}
\lamdd{j} \rp \nonr \\
&+&Tr\lp6\lamd{q}\lamdd{q}\lamd{i}\lamdd{j}+
2\lamd{i}\lamdd{j}\lamet{q}\lamed{q}+2\ye\yed\lamd{i}\lamdd{j}\rp \nonr \\
&+&2\lp\ydd\lamdt{i}\lames{q}\ye\rp_{jq}
+6\lp\ydd\lamdt{i}\lamds{q}\yd\rp_{jq}\nonr \\
&+&4\lp\lamu{m}\rp_{in}\lp\lamud{m}\rp_{jr}Tr\lp\lamdd{n}\lamd{r}\rp
\nonr \\
&+&4\lp\lamu{m}\lp\ydd\yd+\lamus{q}\lamu{q}\rp\lamus{m}\rp_{ij} \nonr \\
&+&2\lp\lamu{n}\lamus{r}\rp_{ij}\lp2\lp\yud\yu\rp_{nr}+Tr\lp\lamud{n}
\lamu{r}\rp\rp,
\\
-\left(\gamma^{(2){U_j}}_{U_i}\right)_{yukawa} &=& 2\lp \yud\yu\yud\yu
\rp_{ji}
+2\lp\yud\yd\ydd\yu\rp_{ji}+6\lp\yud\yu\rp_{ji}Tr\lp\yu\yud\rp \nonr \\
&+&2\lp\yut\lamdd{q}\lamd{q}\yus\rp_{ij}+4Tr\lp\lamud{j}\lamu{i}\ydd\yd\rp
\nonr \\&+&4Tr\lp\lamud{j}\lamu{i}\lamus{q}\lamu{q}\rp +4\lp\lamud{j}
\lamu{i}\rp_{rn}Tr\lp\lamdd{n}\lamd{r}\rp,\lab{gam2uu}
\\
-\left(\gamma^{(2){H_1}}_{H_1}\right)_{yukawa} &=& Tr\lp 3\ye\yed\ye\yed
+9\ydd\yd\ydd\yd+3\yd\ydd\yu\yud\rp\nonr\\
&+&Tr\lp3\lamd{q}\lamdd{q}\ye\yed-\lame{q}\lamed{q}\ye\yed
-6\ydd\yd\lamud{q}\lamu{q} \right. \nonr \\
&+&\left.3\yd\ydd\lamdd{q}\lamd{q}\rp
+\lp\yed\ye\rp_{rn}Tr\lp\lamed{n}\lame{r}\rp \nonr \\
&+&6\lp\ydd\yd\rp_{rn}Tr\lp\lamdd{n} \lamd{r}\rp, \\
-\left(\gamma^{(2){{\bar H}}}_{{\bar H}}\right)_{yukawa} &=&
Tr\lp9\yu\yud\yu\yud+3\yu\yud\yd\ydd+3\yu\yud\lamdt{q}\lamds{q}\rp \nonr \\
&+& 3\lp\yud\yu\rp_{rn}Tr\lp\lamud{n}\lamu{r}\rp,\lab{gam2hbhb}
\\
-\left(\gamma^{(2){H_1}}_{L_i}\right)_{yukawa} &=&
3\lp\ye\yed\lamd{q}\yds\rp_{iq}+\lp\ye\yed\lamet{q}\yes\rp_{iq}\nonr \\
&+&3\lp\yed\ye\yed\lamet{n}\rp_{ni}
+\lp\lame{n}\yes\rp_{ir}Tr\lp\lamed{n}\lame{r}\rp\nonr \\
&+&\lp\lame{m}\lames{q}\lame{q}\yes\rp_{im}
+3\lp\yed\lamd{q}\lamdd{q}\lamet{m}\rp_{mi} \nonr \\
&+&9\lp\ydd\yd\ydd\lamdt{n}\rp_{ni}
+6\lp\ydd\lamdt{n}\rp_{ri}Tr\lp\lamdd{n}\lamd{r}\rp \nonr \\
&+&6\lp\lamus{q} \lamu{q}\ydd\lamdt{n}\rp_{ni}
+3\lp\ydd\yu\yud\lamdt{m}\rp_{mi} \nonr \\
&+&3\lp\lamd{m}\lamdd{q}\lamd{q}\yds\rp_{im}.
\earr
This completes the renormalization group equations for the Yukawa
couplings at two-loop. Before we discuss applications we briefly consider
the renormalization of the bilinear terms.

\subsection{Bi-Linear Terms}
Following the general equations given in MV the renormalization group
equations for the bilinear terms now including all R-parity violating effects
are given by
\barr
\Dt\mu&=&\mu\left\{ \ggam{H_1}{H_1}+\ggam{H_2}{H_2}\right\}
+\kappa^i\ggam{L_i}{H_1},\\
\Dt\kappa^i&=&\kappa^i\ggam{H_2}{H_2}+\kappa^p\ggam{L_p}{L_i}
+\mu\ggam{H_1}{L_i}. \lab{kappa}
\earr
The anomalous dimensions at two-loop are given in the previous subsections.
As already noted in MV the bi-linear terms do not appear in the equations
for the Yukawa couplings.

\subsection{Discussion}
Equation \eq{kappa} implies that for $\kappa_i=0$ at tree-level for all $i$, a
non-zero $\kappa_i$ is generated via the $\mu$-term. However, as is well known,
if the coefficient of the corresponding soft breaking term equals that of the
superpotential term, then the terms $\kappa_iL_iH_1$ in the superpotential can
be rotated away through a redefinition of the $L_i$ and $H_1$ \cite{hallsuz}.
If we are considering the one-loop or two-loop renormalized Lagrangian then we
must make this rotation after renormalization. Again, there will be no term
$\kappa_iL_iH_1$ in the Lagrangian, the rotation matrix will of course differ.
Thus there is no mixing between $L_i$ and $H_1$ and $\Gamma_{L_i}^{H_1}=0$ is
guaranteed by the relevant counterterms at two-loop. This then also applies to
the Eqs(\ref{eq:ye},\ref{eq:yd},\ref{eq:lame},\ref{eq:lamd}) in Section 2.2.1
and Eq.\eq{kappa} in Section 2.3. If the Lagrangian has additional symmetries
which distingiush between $L_i$ and $H_1$ and allows $L_iH_2$ in $W_{\not R_p}$
then these terms must be retained.

The two-loop renormalization group equations for the Yukawa couplings also
respect several symmetries. If at some scale for example $\lam''_{ijk}=0$ for
all $i,j,k$ then baryon parity, $B_p$, is conserved at this scale. There are
no \bpv-couplings in the theory and thus in perturbation theory no
\bpv-couplings are generated, \ie the RGEs preserve $\lam''_{ijk}=0$.
Analogously lepton parity, once imposed, is also preserved by the RGEs. If at
some scale $\lam_{ijk}=\lam''_{ijk}=0$ for all $i,j,k$ and only one lepton
flavour is violated, \eg $\lam'_{3jk}\not=0$ then this is also true for all
scales, provided $\Gamma_{L_n}^{H_1}=0$. So the lepton number violating
couplings $LQ\Dbar$ and $LL\Ebar$  decouple in this limit. If the
neutrino masses are non-zero then this is no longer true. The electron mass
matrix $\ye$ then contains off-diagonal entries which generate off diagonal
$\Gamma_{E_j}^{E_i},\Gamma_{L_j}^{L_i}$ via the RGEs. But the effects will be
very small and can thus be neglected in most circumstances. However, if we
assume only $\lam'_{111}\not=0$ at some scale then through the quark CKM-mixing
the other terms $\lam'_{1ij}$ will be generated.

Our results agree with MV for the MSSM Yukawa couplings. We also agree
with the one-loop results \cite{prob,rogervern,marcgoi}.

\nsect{Unification}
We now apply our two-loop RG-equations to the questions of unification. We
shall assume as a first approximation that the \rpv-couplings have a similar
hierarchy to the  SM Yukawa couplings and thus only consider one coupling at a
time. The third generation couplings have the weakest bounds \eq{bounds} and
can thus lead to the largest effects. We shall consider the three cases
\beq
{\bf LL\Ebar:}\,\;\lam_{323},\quad{\bf LQ\Dbar:}\,\;\lam'_{333},\quad{\bf \Ubar
\Dbar\Dbar:}\,\;\lam''_{323}.
\lab{cases}
\eeq
We assume that in each case the respective operator decouples from the other
\rpv-operators whose couplings we set to zero. Is this approximation
consistent, \ie is it stable under the RGEs? First consider the $LL\Ebar$
operator. The coupling $\lam_{323}$ only violates $L_
\mu$ and could thus also generate non-zero $\lam_{121}$. However, $\lam_{121}$
decouples from $\lam_{323}$ if it is the only non-vanishing operator at some
scale. The other $LL\Ebar$ operators are protected by global $L_e$ and $L_
\tau$. It is thus consistent to only consider a non-zero $\lam_{323}$.

The coupling $\lam'_{333}$ will generate $\lam'_{3ij}\not=0$ in higher order.
However, these terms are proportional to third generation  off-diagonal
CKM-matrix elements which are very small and can be safely ignored. Thus for $
\lam'_{323}$ the decoupling assumption is also good. For $\lam''_{323}$ the
higher order mixing with the couplings $\lam''_{i2k}$ is also very small. The
mixing with $\lam''_{313}$ is Cabbibo suppressed and involves a first
generation Higgs Yukawa coupling and can be neglected. Thus in all three
cases the decoupling is a good assumption. In line with this argument we assume
the following form for the Higgs-Yukawa matrices
\beq
\ye=diag(0,0,\lam_\tau),\quad\yd=diag(0,0,\lam_b),\quad\yu=diag(0,0,\lam_t).
\eeq

In order to determine the scale of unification we numerically solve the
renormalization group equations. Our analysis is analogous to that of Refs
\cite{gunion2} to which we refer for details. We also do not consider GUT
threshold corrections. Our analysis differs in that we restrict ourselves to
three generations but instead add in turn one of the
three \rpv-Yukawa couplings \eq{cases}. Thus we run the full set of equations
including the two-loop correction of the Yukawa couplings to the running of the
gauge couplings \eq{gaugerg}. We use the experimental values $\alpha_{em}^{-1}
(M_Z)=127.9$  and $\sin^2\theta_W(M_Z)=0.2324$ \cite{nirpaul}.
We consider only a single supersymmetry mass scale $M_{SUSY}
=M_Z$. For more realistic spectra $M_{SUSY}$ can be considered an effective
mass scale which enters in the RGEs \cite{nirpaul,marcelacarlos}. The
supersymmetric mass spectrum can be highly non-uniform with the masses
typically larger than $M_{SUSY}$. We then iteratively determine the
value of $\alpha_3(M_Z)$ and thus also the unification mass scale $M_U$ and the
coupling at the unification scale $\alpha_U$ \cite{gunion2}. We have chosen a
running top mass of $m_t(m_t)=165\gev$ which corresponds to a pole mass $m_t=
175\gev$ in agreement with the discovery at the Tevatron \cite{cdf}. We fix
the running bottom quark mass at $m_b(m_b)=4.25\gev$ \cite{gasserleutwyler}.

We solve the RGEs for different values of the \rpv-coupling at the weak
scale, starting from zero. The maximal value we consider is where the running
coupling reaches the perturbative limit at the unification scale. So we
require
\beq
\frac{\lam_{\not R_p}^2(M_U)}{4\pi}<1
\lab{pertlim}
\eeq
or $\lam_{\not R_p}(M_U)<3.5$.

One of the encouraging aspects of supersymmetric grand unified theories is the
possibility of bottom-tau unification at $M_U$. Similar to gauge coupling
unification this is not possible in standard GUTs. There has been a large
interest in the literature \cite{btauguys,fixedpoint1,ohman,marcelacarlos,
nirpaul2} in the
restrictions on the unification scenario from bottom-tau unification. Requiring
bottom-tau unification leads to a strict relation between the running top quark
mass and $\tan\beta$. For the experimental value of $m_t$ \cite{cdf}, $\tan
\beta$ is predicted to be very close to 1.5 or around 60 \cite{figmarcela}. We
are interested in how the effects of \rpv\ can relax this strict relation and
allow a larger range of $\tan\beta$. As a model scenario we consider $\tan\beta
=5$ which is well away from the solutions in the MSSM. We then investigate the
possibilities for a top fix-point solution as well as bottom-tau unification
including the \rpv-effects. We do not consider the other GUT mass ratios.

\subsection{Gauge Coupling Unification}
Before determining the effects of $\lam_{\not R_p}\not=0$ we would also like
to consider the effect from the non-vanishing top quark Yukawa coupling.
Thus we first determine the unification values for $\lam_t=0$ and $\lam_{\not
R_p}=0$. We obtain
\beq
\alpha_3(M_Z)=0.128, \quad M_U=2.98\,10^{16}\gev,\quad \alpha_U=0.043.
\lab{unifvalues}
\eeq
In order to discuss the effects of the non-zero Yukawa couplings we consider
the unification parameters as functions of the weak scale values of $\lam_t$
and $\lam_{\not R_p}$
\beq
\alpha_s(M_Z,\lam_t,\lam_{\not R_p}),\quad
M_U(\lam_t,\lam_{\not R_p}),\quad\alpha_U(\lam_t,\lam_{\not R_p}).
\eeq
and define the ratios
\barr
R_{\alpha_3}(\lam_t,\lam_{\not R_p})&=&\frac{\alpha_3(M_Z,\lam_t,\lam_{\not
R_p})}{\alpha_3(M_Z,0,0)},\nonr\\
R_{M_U}(\lam_t,\lam_{\not R_p})&=&\frac{M_U(\lam_t,\lam_{\not R_p})}{M_U(0,0)},
\lab{ratios}\\
R_{\alpha_U}(\lam_t,\lam_{\not R_p})&=&\frac{\alpha_U(\lam_t,\lam_{\not R_p})}
{\alpha_U(0,0)}.\nonr
\earr
We first consider $\lam_{\not R_p}=0$ and fix $\lam_t(M_Z)=0.44$ so that $m_t
(m_t)=165\gev$ ($\tan\beta=5$). This shifts the ratios \eq{ratios} as is shown
in Figs 1a, 2a, and 3a, $R_i(\lam_t,0)\not=1$. It can be read off for
$(\Lambda_{E^3})_{23}(M_Z)=0$, $(\Lambda_{D^3})_{33}(M_Z)=0$, or $(\Lambda_{U^
3})_{23}(M_Z)=0$ respectively. The shift is of
course identical in the three plots, note however the different scales for $R_i
$. Due to the large top Yukawa $\alpha_s(M_Z)$ and $\alpha_U$ are lowered only
by about $1\%$. The unification scale is lowered by about $1.5\%$.

Next we turn on the \rpv-couplings. In Figures 1b, 2b, and 3b we can read off
the value of the \rpv-coupling at the unification scale as a function of the
coupling at the weak scale. $(\Lambda_{E^3})_{23}(M_U)$ reaches its
perturbative limit \eq{pertlim} for a weak scale value of
\beq
(\Lambda_{E^3})_{23}(M_Z)=0.9.
\eeq
It is worth pointing out that this is the same as the laboratory bound for
slepton masses at $1\tev$! Thus although the laboratory bounds on the
$LL\Ebar$ operators are generally considered to be very strict; for heavy
supersymmetric masses they are no stricter than the perturbative limit.
At this point $(\Lambda_{E^3})_{23}(M_U)$ has run off the plot but it should be
clear how it extrapolates. All other quantities in Figs. 1a,b are plotted up to
$(\Lambda_{E^3})_{23}(M_Z)=0.9$. We have chosen the scaling of the plot so as
to highlight the effects on the other quantities. The perturbative limits for
the other couplings are given by
\barr
(\Lambda_{D^3})_{33}(M_Z)&=& 1.08\\
(\Lambda_{U^3})_{23}(M_Z)&=& 1.14
\earr
The latter value is about $15\%$ higher than the 1-loop value previously
obtained \cite{prob,marcgoi}. This is the same order as the two-loop Yukawa
corrections obtained in \cite{ohman}.

In Figures 1a, 2a, 3a we show how the ratios \eq{ratios} change as we turn on
the \rpv-couplings. For $(\Lambda_{E^3})_{23}(M_Z)$, $\alpha_s$ and $\alpha_U$
are practically unchanged except very close to the perturbative limit. However,
$M_U$ is shifted upwards by up to $10\%$. For $(\Lambda_{E^3})_{23}(M_Z)\approx
\lam_t(M_Z)$ the decrease due to the top quark is off-set by $\lam_{\not R_p}$.
For $(\Lambda_{D^3})_{33}(M_Z)$ the maximum combined shift in $\alpha_s(M_Z)$
is a decrease of $4\%$ giving a value of $\alpha_s(M_Z)=0.123$, in better
agreement with the data \cite{alphas}. $\alpha_U$ is decreased slightly.
However, $M_U$ is decreased by up to $20\%$. This effect
is significantly beyond the effect due to the top quark Yukawa coupling.
For $(\Lambda_{U^3})_{23}(M_Z)$, $\alpha_U$ remains practically unchanged.
$\alpha_s$ now has an overall increase of about $5\%$ at the perturbative limit
corresponding to a value of $\alpha_s(M_Z)=0.134$ in disagreement with the
experimental value. $M_U$ is raised by up to $20\%$.

Thus we find $\alpha_U$ essentially unchanged by \rpv-effects. $M_U$ can change
either way by up to $20\%$. If we compare this with other effects considered in
Ref. \cite{nirpaul} we find it of the same order as the uncertainty due to the
top quark Yukawa coupling or the effects of possible non-renormalizable
operators at beyond the GUT scale. The effect is much smaller
than that due to GUT-scale threshold corrections or weak-scale supersymmetric
threshold corrections. It is thus much too small an effect to accomodate string
unification. The strong coupling can also change either way by up to $5\%$. A
decrease is favoured by the data and is welcome in supersymmetric unification.
The effect of the \rpv-couplings on $\alpha_s(M_Z)$ is of the same order as the
effects due to the top-quark Yukawa coupling, GUT-scale threshold effects and
high-scale non-renormalizable operators \cite{nirpaul}.

\subsection{b-$\tau$ Unification}
In order to study the unification of the bottom and $\tau$ Yukawa couplings
$\lam_b,\,\lam_\tau$ at $M_U$ we define the ratio
\beq
R_{b/\tau}(M_U)=\frac{\lam_b(M_U,\lam_{\not R_p})}{\lam_\tau(M_U,\lam_{\not
R_p})}.
\eeq
For $\lam_{\not R_p}=0$, $\tan\beta=5$ we have
\beq
R_{b/\tau}(M_U)=0.74.
\eeq
Thus including the top-quark effects but before turning on the \rpv-coupling
we are well away from the bottom-tau unification solution $R_{b/\tau}(M_U)=1.$
Recall that the uncertainties due to the bottom quark mass are small for small
$\tan\beta$. Now we consider the corrections due to the \rpv-couplings. The
one-loop RGE for $R_{b/\tau}(t)$ is given by
\beq
\frac{dR_{b/\tau}(t)}{dt}=R_{b/\tau}(t)\left[ \lam_t^2+3\lam_b^2-3\lam_\tau^2
-3\lam_{323}^2+2\lam^{''2}_{323}+\frac{4}{3}g_Y^2-\frac{16}{3}g_3^2\right].
\eeq
At one-loop the evolution of $R_{b/\tau}$ is unaffected by the
operator $L_3Q_3\Dbar_3$. The slight rise which we observe in Fig 2b is a
two-loop effect which is very small. The leading dependence of $R_{b/\tau}$
on $\lam_{323}$ has a negative sign and as we see in the two-loop result shown
in Fig. 1b $R_{b/\tau}$ drops significantly. At the perturbative limit it has
dropped by a factor of three and \rpv\  becomes a dominant effect on the
evolution of $R_{b/\tau}$. This is important for the range of $\tan\beta$ which
leads to bottom-tau unification. In the MSSM $R_{b/\tau}$ is too large for
$\tan\beta\lsim1.5$ or $\,\,\gsim60$ \cite{figmarcela}. Including a non-zero
operator $\lam_{323}$ strongly reduces $R_{b/\tau}$ and thus can lead to
bottom-tau unification outside the previous regime.

For $\lam''_{323}\not=0$ there is an additional {\it positive} contribution in
the evolution of $R_{b/\tau}(t)$. The full two-loop result shows a clear rise
in $R_{b/\tau}(t)$ as a function of $\lam''_{323}$ in Fig. 3b. The maximum
increase at the perturbative limit is by $60\%$. For $\lam''_{323}(M_Z)=1.07$
bottom-tau unification is restored! This is quite remarkable. Even though
\rpv-couplings are usually expected to lead to only small effects they can have
a significant impact on our understanding of Yukawa-unification. Recall, that
grand unification {\it is} possible in \rpv-theories \cite{hallsuz}.

{}From the structure of the couplings it should be clear that for example for
$\lam'_{233}$ we get an increase in $R_{b/\tau}(M_U)$ as well since at one-loop
it does not couple to $\lam_\tau$. Similarly, $\lam'_{322}$ decreases $R_{b/
\tau}$. This leads to further bottom-tau unification points.

\subsection{Fixed Point Structure of the Top Yukawa}
We would next like to discuss the effects of the \rpv-couplings on $\lam_t(M_U)
$. In particular we are interested in finding fix-point structures. These are
defined by zero derivative at $M_Z$. For the MSSM parameter-point we have
chosen, $\tan\beta=5$, we obtain
\beq
\lam_t(M_U)=0.45,\quad {\rm for} \lam_{\not R_p}=0.
\eeq
This is indicated in Figs 1b, 2b, and 3b. The one-loop evolution equation of
$\lam_t$ is given by
\beq
\frac{d\lam_t}{dt}=\lam_t\left[6\lam_t^2+\lam_b^2+2\lam^{''2}_{323}+\lam^{'2}
_{333}-(\frac{13}{15}g_Y^2+3g_2^2+\frac{16}{3}g_3^2)\right],
\eeq
{}from which we can read off the dominant effects. At one-loop, the evolution
of $\lam_t$ is independent of the $LL\Ebar$ operators. If we look at equation
\ref{eq:rgyup}, as well as the equations for the two-loop anomalous dimensions
$\Gamma_{Q_3}^{Q_3}$, $\Gamma_{U_3}^{U_3}$, and $\Gamma_{\Hbar}^{\Hbar}$
we see that for $\lam'_{ijk}=0$, for all $i,j,k$, there is also no dependence
on $LL\Ebar$ at two-loops. Thus $\lam_t(M_U)$ is independent of $\lam_{323}
(M_Z)$ at two-loop, as we see in Fig. 1b.

For $L_3Q_3\Dbar_3$ we see that $\lam_t(M_U)$ should grow with $\lam'_{333}$.
The full two-loop effect is shown in Fig. 2b. At the perturbative limit
$\lam_t$ is large but clearly unequal to 1. The increase is by about $60\%$.
The derivative at $M_Z$ does not vanish and we do not observe fix-point
structure for $\lam_t$ or $\lam'_{333}$. For lower values of $\tan\beta$
but well above 1.5 new fixed point solutions exist.

For $\Ubar_3\Dbar_2\Dbar_3$ we have a larger positive coefficient than in the
previous case and as we see in Fig 3b $\lam_t$ rises more quickly as we
approach the perturbative limit. For $\lam''_{323}(M_Z)\approx 1$, $\lam_t$
has the fix-point structure, \ie
\beq
6\lam_t^2(M_Z)+2\lam^{''2}_{323}(M_Z)\approx\frac{16}{3}4\pi^2\alpha_3(M_Z).
\eeq
This is roughly the same point where $R_{b/\tau}\approx1$. Thus for $\Ubar_3
\Dbar_2\Dbar_3$ bottom-tau unification also corresponds to top IR fix-point
structure, as in the MSSM \cite{ir,topfix,ohman}. This is all quite remarkable.
The \rpv-couplings can have significant affects on the entire Yukawa
unification picture. From the one-loop RGE for $\lam''_{323}$ we can see that
we are well away from its infra-red fixed-point.

\nsect{Conclusions}
We have argued that \rpv\ is theoretically on equal footing with conserved
$R_p$. Since it can be realized in grand unified theories it is relevant for
unification. We then first determined the complete two-loop renormalization
group equations for the dimensionless couplings of the unbroken supersymmetric
Standard Model. It is only at two-loop that Yukawa couplings affect the running
of the gauge coupling constants. We then considered three models of \rpv. We
have added to the MSSM in turn the three Yukawa operators $L_3L_2\Ebar_3$, $L_3
Q_3\Dbar_3$, and $\Ubar_3\Dbar_2\Dbar_3$. We considered their effects on
various aspects of the perturbative unification scenario. We have focused on
qualitative effects. A detailed search for a preferred model is beyond the
scope of this paper. We found several important effects. The unification scale
is shifted by upto $\pm20\%$. This is comprable to some threshold effects but
insufficient for string unification. $\alpha_s(M_Z)$ can be changed at most by
$\pm5\%$. The reduction which is favoured by the data is obtained close to the
perturbative limit of $\lam'_{333}$. We have obtained the two-loop limit from
perturbative unification for all three operators. For $\lam_{323}$ it is
equivalent to the laboratory bound for a slepton mass of $1\tev$. For $\lam''_
{323}$ the two-loop limit is $15\%$ weaker than the one-loop limit previously
obtained. For bottom-tau unification we have found significant affects. For
$L_3L_2\Ebar_3$ bottom-tau unification can be obtained for values of $\tan
\beta<1.5$. For $\Ubar_3\Dbar_2\Dbar_3$ we found a new point of bottom-tau
unification at $\tan\beta=5$. It is remarkable that this point also shows
top Yukawa infra-red fixed point structure but no $\lam''_{323}$ fixed point
structure.

{\bf Acknowledgements}\newline
We thank Vernon Barger, Probir Roy and Marc Sher for helpful converstations.
H.P. would like to thank J. Gunion for a fruitful collaboration which
contibuted to this work. This work was partially supported by the National
Science Foundation under grant no. PHY94-07194.

\nappend{Appendix}
We consider a group $G$ with representation matrices ${\bf t}^A\equiv
({\bf t})^{Aj}_i$. Then the quadratic Casimir $C(R)$ of a representation
$R$ is defined by
\beq
({\bf t}^A{\bf t}^B)^j_i= C(R) \delta^j_i.
\eeq
For $SU(3)$ triplets $q$ and for $SU(2)_L$ doublets $L$ we have
\beq
C_{SU(3)}(q)=\frac{4}{3}, \quad C_{SU(2)}(L)=\frac{3}{4}.
\eeq
For $U(1)_Y$ we have
\beq
C(f)=\frac{3}{5} Y^2(f),
\eeq
where $Y(f)$ is the hypercharge of the field $f$. The factor $3/5$
is the grand unified normalization.

For the adjoint representation of the group
of dimension $d(G)$ we have
\beq
C(G)\delta^{AB}=f^{ACD}f^{BCD},
\eeq
where $f^{ABC}$ are the structure constants. Specifically for the
groups we investigate
\beq
C(SU(3)_C)=3,\quad C(SU(2)_L)=2,\quad C(U(1)_Y)= 0,
\eeq
and $C(SU(N))=N$. The Dynkin index is defined by
\beq
Tr_R({\bf t}^A{\bf t}^B)\equiv S(R)\delta^{AB}.
\eeq
For the respective fundamental representations $f$ we obtain
\barr
SU(3),\, SU(2):\quad &&S(f)=\half,\\
U(1)_Y:\quad&&S(f)=\frac{3}{5}Y^2(f),
\earr
where we have inserted the GUT normalization for $U(1)_Y$.

The coefficients in the two-loop running of the gauge couplings
\eq{gaugerg} are given by \cite{bjorkjones}
\barr
B_a^{(1)}&=&(\frac{33}{5},1,-3),\\
B_{ab}^{(2)}&=&\left( \begin{array}{ccc}
199/25 & 27/5 & 88/5 \\
9/5 & 25 & 24 \\
11/5 & 9 & 14
\end{array}
\right),\\
C_{a}^{u,d,e}&=&\left( \begin{array}{ccc}
26/5 & 14/5 & 18/5 \\
6 & 6 & 2 \\
4 & 4 & 0
\end{array}
\right).
\earr

\noindent {\bf Figure 1:} We have fixed the values $\tan\beta=5$,
$m_b(m_b)=4.25\gev$ and $m_t(m_t)=165\gev$. For a vanishing top-quark Yukawa
coupling and a vanishing \rpv-coupling we obtain the unification values shown
in Fig (1a). In (1a) we then show the ratios $R_{M_U}$, $R_{\alpha_U}$
and $R_{\alpha_3}$ at the unification scale as a function of
$(\Lambda_{E^3})_{23}(M_Z)$. In Fig. (1b) we show the values of $R_{b/\tau}$,
$\lam_t$ and $(\Lambda_{E^3})_{23}$ at the unification scale as a function
of $(\Lambda_{E^3})_{23}(M_Z)$. $(\Lambda_{E^3})_{23}(M_U)$ reaches its
perturbative limit for $(\Lambda_{E^3})_{23}(M_Z)=0.9$. At this point it is
outside the plotted region. The plotted region in Figs (a) and (b) is chosen
so as to highlight the evolution of the other quantities.

\noindent {\bf Figure 2:} The same as Figure 1 just for $(\Lambda_{D^3})_{33}$
except the perturbative limit is reached for $(\Lambda_{D^3})_{33}(M_Z)=1.08$.

\noindent {\bf Figure 3:} The same as Figure 1 just for $(\Lambda_{U^3})_{23}$
except the perturbative limit is reached for $(\Lambda_{U^3})_{23}(M_Z)=1.14$.

\end{document}